\documentclass{raa}
\usepackage{graphicx,times}
\usepackage{natbib}
\usepackage{amssymb,amsmath}
\usepackage[pagebackref=true]{hyperref}
\bibpunct{(}{)}{;}{a}{}{,}

\usepackage{lineno}
\graphicspath{{./}{figures/}}
\newcommand{\ha}{H\(_\alpha\)}
\newcommand{\hb}{H\(_\beta\)}
\newcommand{\logt}{\(\mathrm{log}~\tau_{~500nm}\)}

\newcommand{\sm}{$\sim$}
\newcommand{\jw}[1]{\textcolor{blue}{ #1}}
\makeatletter
\renewcommand\sout{\bgroup\markoverwith{\textcolor{red}{\rule[0.5ex]{2pt}{1.5pt}}}\ULon}
\makeatother

\begin{document}
	\title{Magnetic Field Measurements in the Solar Chromosphere Using the \hb\ 4861\AA~Line I: Forward Modeling Based on 1D Models}
	\volnopage{ {\bf 20XX} Vol.\ {\bf X} No. {\bf XX}, 000--000}
	\setcounter{page}{1}
	
	\author{
		Jiasheng Wang\inst{1},
		Wenxian Li\inst{2,*},
		Xianyong Bai\inst{2,3},
		Yingzi Sun\inst{2},
		Yuanyong Deng\inst{2,3},
		\and Jiaben Lin\inst{2}
		\footnotetext{$*$Corresponding Author}
	}
	
	\institute{
		School of Earth and Space Sciences, Peking University, Beijing 100871, China\\
		\and
		State Key Laboratory of Solar Activity and Space Weather, National Astronomical Observatories, Chinese Academy of Sciences, Beijing 100101, China; {\it wxli@nao.cas.cn}\\
		\and
		School of Astronomy and Space Science, University of Chinese Academy of Sciences, Beijing, China}

	\abstract{
		The chromosphere is a complex solar atmosphere that hosts a variety of transients and transports significant free energy to heat the corona.
		However, due to the limited sensitivity of polarization measurement and the influence of spectral line broadening, the basic magnetic field configuration in the chromosphere has not yet been fully revealed to correspond to the observed phenomena. In this work, we investigated the validity and application of the magnetic field inversion method for the \hb\ 4861 \AA\ spectral line with non-local thermodynamic equilibrium (non-LTE) approximations. The formation height of the \hb\ line center in the chromosphere is 1100--1300~km (\logt  = [$-$4.86, $-$5.00]) in quiet Sun and 300 -- 410~km (\logt = [$-$2.41,$-$4.57]) in sunspots. We generated synthetic spectra by incorporating magnetic fields into semi-empirical FAL models for quiet Sun and sunspots, and then performed inversions to obtain the magnetic fields, which were then compared with the magnetic fields in the models. 
		In addition, we evaluated the accuracy of the magnetic fields obtained using the weak-field approximation (WFA) and the impact of using these WFA results as the initial guess model for non-LTE inversion on the final results. 
		Our work validates the effectiveness of the inversion method for the measurement of line-of-sight (LOS) magnetic field components, which significantly improved the accuracy in both weak field (0 -- 500~G) and strong field ($>$2000~G) regions, while maintaining accuracy in the intermediate field range of 500 -- 2000~G. This demonstrates that the inversion techniques we employed are capable of resolving Zeeman-sensitive spectral lines in the chromosphere, which can be applied to the \hb\ observational data from the new generation Solar Full-disk Multi-layer Magnetograph at GanYu Solar Station to provide full disk chromospheric magnetic field information.
	}
	
	\authorrunning{J. Wang et al. }            
	\titlerunning{Chromospheric Field Inversion with 1D NLTE}  
	\maketitle
	
	\section{Introduction}\label{sec:intro}
	
	The solar chromosphere is a highly dynamic atmospheric layer, hosting numerous transient phenomena and playing a pivotal role in the upward transport of magnetic energy that heats the corona. Mechanisms such as magnetoacoustic wave propagation, nanoflares, and spicules have been proposed to contribute to this energy transport, supported by extensive observational evidence \citep[e.g.,][]{2007Sci...318.1574D, 2014Sci...346D.315D, 2014Sci...346A.315T, 2019Sci...366..890S}. Magnetic reconnection and wave dissipation processes occurring across various spatial and temporal scales significantly alter the chromospheric magnetic structure, both in active regions and the quiet Sun environment \citep[e.g.,][]{2017LRSP...14....2B, 2025arXiv250312235C}.
	
	The evolution of the solar magnetic field governs the energy budget and onset of dynamic events, including solar flares, ultraviolet (UV) bursts, and mini-filament eruptions \citep[e.g.,][]{2016ApJ...824...96T, 2018ApJ...854...92T, 2024ApJ...974..123W}. 
	Magnetic fields in solar atmosphere have been measured using numerous methods, such as spectro-polarimetry, magnetic-field-induced transition, seismological measurements of Alfv\'enic waves \citep[refer to][for a review]{yang2022}, also see \cite[][]{2023RAA....23b2001C, 2025RAA....25a5010G}. 
	However, high-resolution measurements of the magnetic field above the photosphere are still very difficult, although significant advances have been made recently \citep[e.g.,][]{2020ScChE..63.2357Y, 2020Sci...369..694Y, 2024Sci...386...76Y, 2020Sci...367..278F, 2024SciA...10.1604S, 2025arXiv250808970C}.
	The low density environment in the chromosphere results in much less frequent particle collisions and cause the deviation of population of energy states in atoms to deviate from LTE. Therefore, solving the full RTE coupled with statistical equilibrium (SE) equations is essential for realistic spectral synthesis and inversion of chromospheric lines (\cite{2012A&A...543A..34D}).
	
	Linear polarization in these lines is influenced by the asymmetry of electron scattering, both in the solar atmosphere and in stellar environments \citep{1974MNRAS.167P..27C}. 
	Therefore, despite their importance, measurements of the chromospheric magnetic field remain challenging due to complex radiative transfer conditions and limited sensitivity of polarization diagnostics. 
	In recent years, studies employing lines such as UV lines of Mg II h\&k \citep{2024ApJ...975..110L}, the visible line of He I D3 \citep{2025A&A...696A.109E}, and IR lines of Ca II triplet \citep{2017A&A...599A.133A} and He I \citep{2021ApJ...921...39A} have advanced our understandings of spectropolarimetric inversions of chromospheric lines by accounting for radiative transfer and non-LTE effects \citep[see the review of][]{2017SSRv..210..109D,araa2022}. 
	However, comprehensive vector field reconstructions are still constrained by instrumental limitations and the complex interplay between magnetic fields, velocity fields, and non-LTE effects.
	
	Among the diverse chromospheric spectral lines, hydrogen lines hold significant diagnostic potential. In flare active regions, electron beams significantly impact hydrogen line profiles through nonthermal processes that are highly sensitive to the local plasma distribution \citep{1993SoPh..143..259Z}. These electron beams retain higher densities than expected under full ionization conditions, due to slower decreases in ionization levels. This enhances the rate of nonthermal excitation and ionization of hydrogen atoms \citep{1992KPCB....8c..34Z}.
	Variations in the intensity of line cores and wings in emergent \hb\ radiation are closely related to atmospheric heating, electron beam input, and collisional energy loss \citep{2009A&A...499..923K}. Enhanced emission at secondary formation heights in the chromosphere, frequently observed as central reversals in line profiles, are indicative of nonthermal collisional effects \citep[e.g.,][]{1990ApJ...348..333C,1992ApJ...401..761D,2015ApJ...813..125K}.
	While the intensity ratios of hydrogen Lyman series lines are largely insensitive to magnetic field strength \citep{2009A&A...504..239T}, \citet{2017ApJ...850...36C} used simultaneous observations of Balmer \ha\ and \hb\ lines to explore the correlation between the \ha/\hb\ ratio and energy flux. Their findings suggest differential energy deposition between the respective formation heights in the chromosphere. Furthermore, spectral observations of hydrogen Balmer lines (\ha, \hb) reveal significant intensity excesses during flare phases, attributed to sudden energy injections \citep{2014ApJ...794L..23H, 2015ApJ...813..125K, 2017ApJ...850...36C}.
	Dynamic processes in the solar chromosphere temporarily redistribute atomic populations through both thermal and nonthermal radiation. The onset of solar flares has been observed to coincide with impulsive Balmer continuum emission arising from hydrogen recombination \citep{2014ApJ...794L..23H}.
	While hydrogen spectral lines are widely used as diagnostics for geometric and dynamic properties of the solar atmosphere, their potential for magnetic field measurements through polarization signals remains relatively underutilized and technically challenging \jw{\citep{2024ApJ...971...30M,2025ApJ...987L..39M}}.
	
	In this study, we perform magnetic field inversion using polarized \hb\ (4861 \AA) spectral lines, employing a non-LTE inversion technique. Our objective is to establish a balanced method that optimizes both computational efficiency and accuracy in reconstructing magnetic field vectors from chromospheric polarization data.
	We evaluate the applicability and performance of this non-LTE based inversion approach for diagnosing the solar chromosphere using \hb\, contributing to improved methods for interpreting solar magnetic structures under complex atmospheric conditions. 
	Section \ref{sec:spectra} gives an overview of the \hb\ line.
	Section \ref{sec:methods} introduces the inversion methods and model atmospheres employed in the paper for interpreting the chromospheric magnetic fields. Section \ref{sec:results} presents inversion results from 1D model simulation with different magnetic field configurations. Finally, we conclude the validation of inferring chromospheric magnetic field using formal solution under non conditions, and discuss the limitations of 1D model application.

	\section{\hb\ 4861~\AA~line} \label{sec:spectra}
	
	\begin{figure}[ht!]
		\centering
		\includegraphics[width=\linewidth]{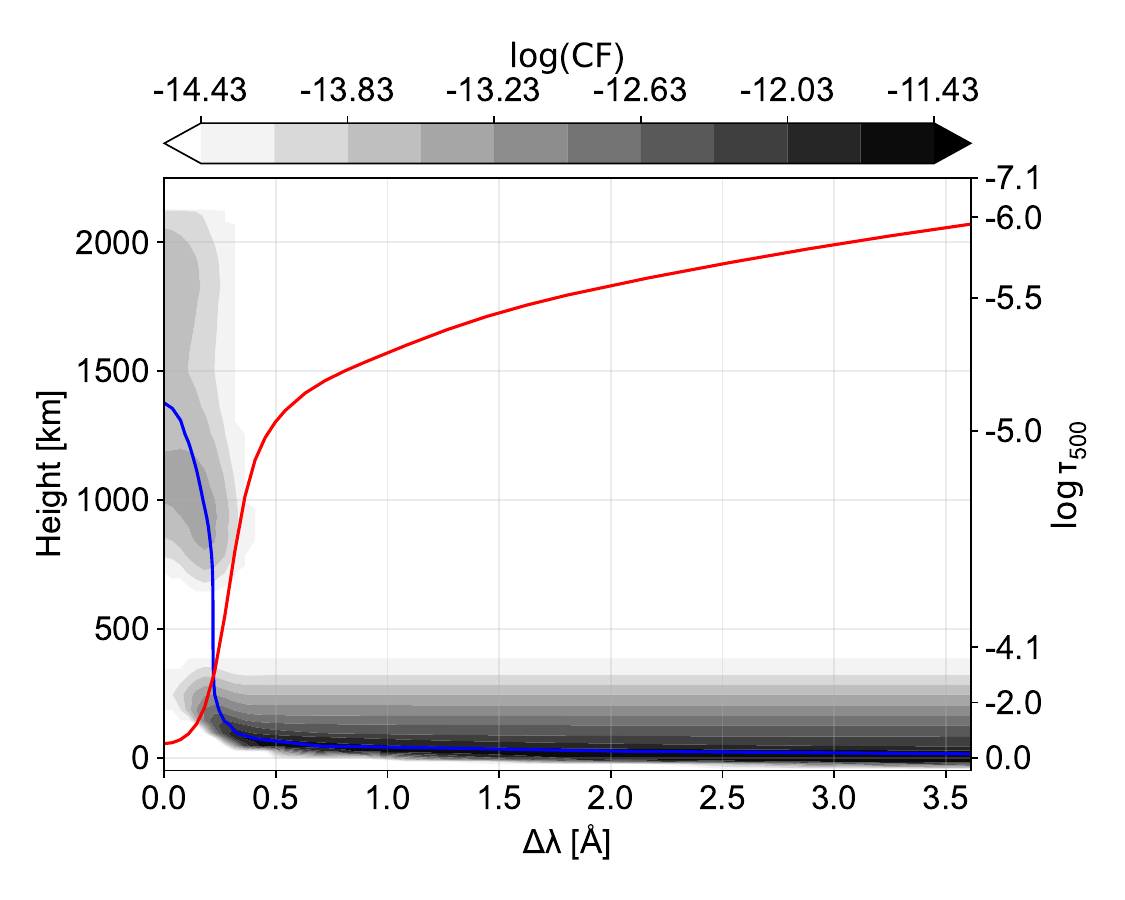}
		\caption{Contribution function (CF) of Stokes I (red curve) of synthetic in \hb\ 4861.3~\AA\ for FAL-C model atmosphere at wavelength $\Delta \lambda$ = [0, 3.6] \AA~from \hb\ line center. The blue curve indicates optical depth \logt = 1 in \hb\ line.}
		\label{fig:cf}		
	\end{figure}

	The \hb\ line forms across a broad range of the plasma-$\beta$ regime in the solar atmosphere. As shown in Figure \ref{fig:cf}, the contribution function of the emergent Stokes I profile exhibits two distinct peaks: one at the line wing (wavelength offset greater than 0.1 \AA\ from the line core) and another at the line core (wavelength offset less than 0.1 \AA). The peak contribution height corresponding to the line wing is 300~km, while that of the line core is 1000 -- 1500~km above the solar surface (corresponding to \logt~= 1, where \logt~is the logarithm of the optical depth scale at 5000 \AA), indicating dual formation heights that span both the lower photosphere and the middle chromosphere.
	
	An early effort to model the \hb\ polarization was conducted by \citet{1986AcASn..27..217Z}, who solved the Unno-Beckers RTEs using the VAL model of the quiet-Sun chromosphere to derive numerical solutions for the Stokes parameters of the \hb\ line. 
	\cite{1986A&A...156...79B, 1986A&A...156...90B} found that the \hb\ line is depolarized not only by the Hanle effect but also by collisions with electrons and protons in low-density medium ({\it n} $\sim 10^{10}$ cm$^{-3}$) and thus the linear polarization measurements in \hb\ line provided a method for determining both the magnetic field and the electron density in quiescent prominences.
	However, in high-density plasma environments such as prominences or filaments (with electron densities $>10^{11}~\mathrm{cm}^{-3}$), collisional depolarization becomes dominant, effectively suppressing any measurable polarization in the \hb\ line \citep{1986A&A...156...79B}. 
	Further investigation by \citet{2002ChJAA...2...71Q} confirmed that the \hb\ line forms at two distinct heights, one in the photosphere and one in the middle chromosphere, by analyzing the contribution function under quiet-Sun conditions. This dual formation height was also demonstrated by \citet{2000SoPh..196..269Z}, showing clear peaks in the contribution to the emergent Stokes parameters, corresponding to changes in local atmospheric conditions.
	
	\cite{2000SoPh..194...19Z} found similarity of quiet Sun magnetic structure between the photosphere and chromosphere by comparing Fe I 5324~\AA\ and \hb\ magnetograms, especially the vertical magnetic flux is conserved and maintains its magnetic feature configuration (topology) at different heights.
	The twisted magnetic configuration was interpreted from the observation of \hb\ magnetograms in quiet Sun prominences \citep{2003ChJAA...3...87B}. 
	\cite{2020SCPMA..6319611Z} found that in the model umbral atmosphere, the departure coefficients characterizing the non-LTE populations of the \hb\ line remain relatively stable from the temperature minimum region to the upper chromosphere.
	Recent study by \cite{2024ApJ...965...15K} have revealed the morphological characteristics of chromospheric fine-scale fibril structures and associated magnetic canopy from the \hb\ filtergrams with VBI deployed at the Daniel K. Inouye Solar Telescope \citep[DKIST;][]{2020SoPh..295..172R} and the \hb\ spectral imaging taken by CHROMIS dual Fabry-Perot interfer-ometer installed at the SST \citep{2003SPIE.4853..341S,2017psio.confE..85S}.

	\section{Methodology} \label{sec:methods}
	\subsection{Spectrum synthesis} \label{subsec:synthesis}
	
	In this work, we adopted the semi-empirical one-dimensional atmosphere FAL models  encompassing the photosphere, chromosphere, and transition region \citep{2006ApJ...639..441F, 2007ApJ...667.1243F, 2009ApJ...707..482F}. 
	FAL models provide sufficient accuracy of temperature structure with modification around temperature minimum region between photosphere and chromosphere \citep{1985cdm..proc...67A,1986ApJ...306..284M}.
	The polarization profiles of \hb\ line are generated using FAL-C atmosphere for quiet Sun and FAL-S variant for sunspot \citep{1990ApJ...355..700F,1991ApJ...377..712F,1993ApJ...406..319F}, with ad hoc magnetic fields of varying strengths and configurations. 
	
	\begin{figure}[ht!]
		\centering
		\includegraphics[width=\linewidth]{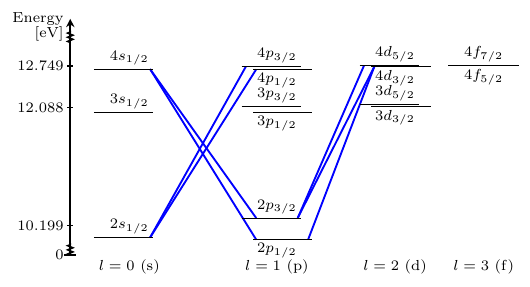}
		\caption{Grotrian diagram of hydrogen Balmer series including first two ordered lines. The blue lines indicate 7 allowed E1 transitions correspond to \hb. The y axis indicates relative level energies  (term value in eV) of each state.}
		\label{fig:grotrian}		
	\end{figure}
	
	In the simulation, we assume non-LTE condition of background hydrogen atoms with fixed departure coefficients, and compute 9 levels of energy transitions of hyperlines including Lyman, Balmer, and Paschen series.
	The required atomic level populations are determined by solving SE equations with the numerical radiative transfer code Rybicki-Hummer (RH) \citep{2001ApJ...557..389U}, which are then used to calculate the departure coefficients of lower and upper levels of permitted transitions.
	With knowledge of the formation of the \hb\ lines and given coefficients of atomic states, the forward calculation was carried out by solving the numerical integration of the RTEs iteratively.	
	\hb\ has a combination of 7 allowed E1 transitions between fine structure state of n=4 to n=2 (see Grotrian diagram of Fig. \ref{fig:grotrian}). The redistributed populations across transitions of fine structures for all $njl$-states were kept constant in the simulation to maintain field-free approximation. 
	Collision broadening of the profiles is calculated using Anstee, Barklem and O'Mara's line broadening parameters \citep{1995MNRAS.276..859A, 1997MNRAS.290..102B} for collisions with neutral hydrogen atoms, the same as those used in \cite{2021A&A...652A.161Q}. Collisional rate coefficients are adapted from \cite{1972ApJ...174..227J}. 
	
	When performing synthesis with RH code for emergent Stokes profiles, the \hb\ absorption profiles were treated with PRD effect.
	The synthetic emergent intensity and polarization profiles were computed along the heliocentric angle $\mu$=1. The spectral line was computed over a 3~\AA\ wavelength range (1.5~\AA\ from each side of the line center). A constant macro-turbulent velocity of 4~km~s$^{-1}$ along atmospheric height was adapted in the model atmosphere.
	
	The Stokes profiles used for inversion are convolved with instrumental broadening, self-broadening, and radiative damping profiles. 
	The instrumental broadening is introduced by convolving filter bandpass of the Solar Full-disk Multi-layer Magnetograph (SFMM) of the Chinese Meridian Project Phase II \citep{ziwuerqi}, which is 85~m\AA. As a result, the instrument broadening that account for Doppler width offset is 10~m\AA~comparing to synthesis. 
	In addition, to mimic real spectropolarimetric observations, we added a noise level of $\sigma_{n}=1.5\times10^{-3}$ to the synthetic Stokes profiles relative to the continuum intensity.
	
	\subsection{Methods for inferring the magnetic fields}
	To retrieve magnetic field information from the chromospheric \hb\ 4861.36 \AA~ line, we employ both the Weak-Field Approximation \citep[WFA; e.g.,][]{2017SSRv..210..109D, 2018ApJ...866...89C} and non-LTE inversion techniques \citep[DeSIRe;][]{2022A&A...660A..37R}. These two approaches are totally different in terms of realism and computational complexity. 
	
	\subsubsection{WFA\label{subsec:wfa}}
	
	WFA is widely accepted in inferring magnetic field when Zeeman splitting is much smaller than the typical (Doppler) width of a spectral line. The line widths of chromospheric profiles are large enough to allow estimation with the WFA for stronger field strength than in the photosphere. 
	\cite{2018ApJ...866...89C} investigated the validity and limitations of the WFA when applied to full Stokes Ca II 8542 Å profiles to extract information about the chromospheric magnetic field.
	
	Taking into account first- and second-order perturbation equations in \citet{:2004aa} derivation, line-of-sight (LOS) components of the vector magnetic field are calculated under strong assumptions: transverse component of magnetic field, LOS velocity, and Doppler width are all independent of optical depth ($\tau_c$). 
	The first order perturbation yields: 
	\begin{equation}
		V = - \Delta \lambda_B\overline{g}\cos\theta \frac{\partial I}{\partial \lambda},
		\label{eq:1}
	\end{equation}
	where $ \lambda_B$ is Zeeman splitting, defined as $\lambda_B = 4.67\times10^{-13}\lambda_0$B, and $\overline{g}$ is the effective $Land\acute{e}$ factor. 
	
	The expression of linear polarization is derived from second-order perturbation at line center:
	\begin{equation}
		U = - \frac{1}{4}\Delta \lambda^2_B\overline{G}\sin^2\theta\sin2\phi_B \frac{\partial^2 I}{\partial \lambda^2}, 
		\label{eq:2}
	\end{equation}
	
	\begin{equation}
		Q = - \frac{1}{4}\Delta \lambda^2_B\overline{G}\sin^2\theta\cos2\phi_B \frac{\partial^2 I}{\partial \lambda^2}, 
		\label{eq:3}
	\end{equation}    
	
	and at wavelength $\lambda_w$ away from line center $\lambda_0$: 
	\begin{equation}
		U =  \frac{3}{4}\Delta \lambda^2_B\overline{G}\sin^2\theta\sin2\phi_B \frac{1}{\lambda_w - \lambda_0} \frac{\partial^2 I}{\partial \lambda^2}, 
		\label{eq:4}
	\end{equation}
	
	\begin{equation}
		Q =  \frac{3}{4}\Delta \lambda^2_B\overline{G}\sin^2\theta\cos2\phi_B \frac{1}{\lambda_w - \lambda_0} \frac{\partial^2 I}{\partial \lambda^2}, 
		\label{eq:5}
	\end{equation}
	
	where $\phi_B$ is azimuth angle of {\bf B} with respect to reference direction, and $\overline{G}$ is the $Land\acute{e}$ factor for transverse field. It is specific to the energy levels involved in transitions.
	
	Under the weak-field approximation (WFA), the circular polarization described by Eq.~(\ref{eq:1}) provides a direct measurement of the LOS component of the magnetic field ($B_{LOS} = B\cos\theta$) through the proportionality between Stokes {\it V} and the first derivative of Stokes {\it I}. Combining Eqs.~(\ref{eq:2}) -- (\ref{eq:3}) at line center (or Eqs. (\ref{eq:4}) -- (\ref{eq:5}) at $\lambda_w$), the linear polarization signals provide constraints on the transverse component ($B_{\perp}$) of the magnetic field via their dependence on $B\sin\theta$. Then the total magnetic field strength is reconstructed as:
	\begin{equation}
		B =  \sqrt{B^{2}_{LOS}+B^{2}_{\perp}}. 
		\label{eq:6}
	\end{equation}
	The inclination angle may then be obtained from the ratio $\tan\theta = B_{\perp}/B_{LOS}$, when both components are reliably measured.

\begin{figure}[ht!]
		\centering
		\includegraphics[width=\linewidth]{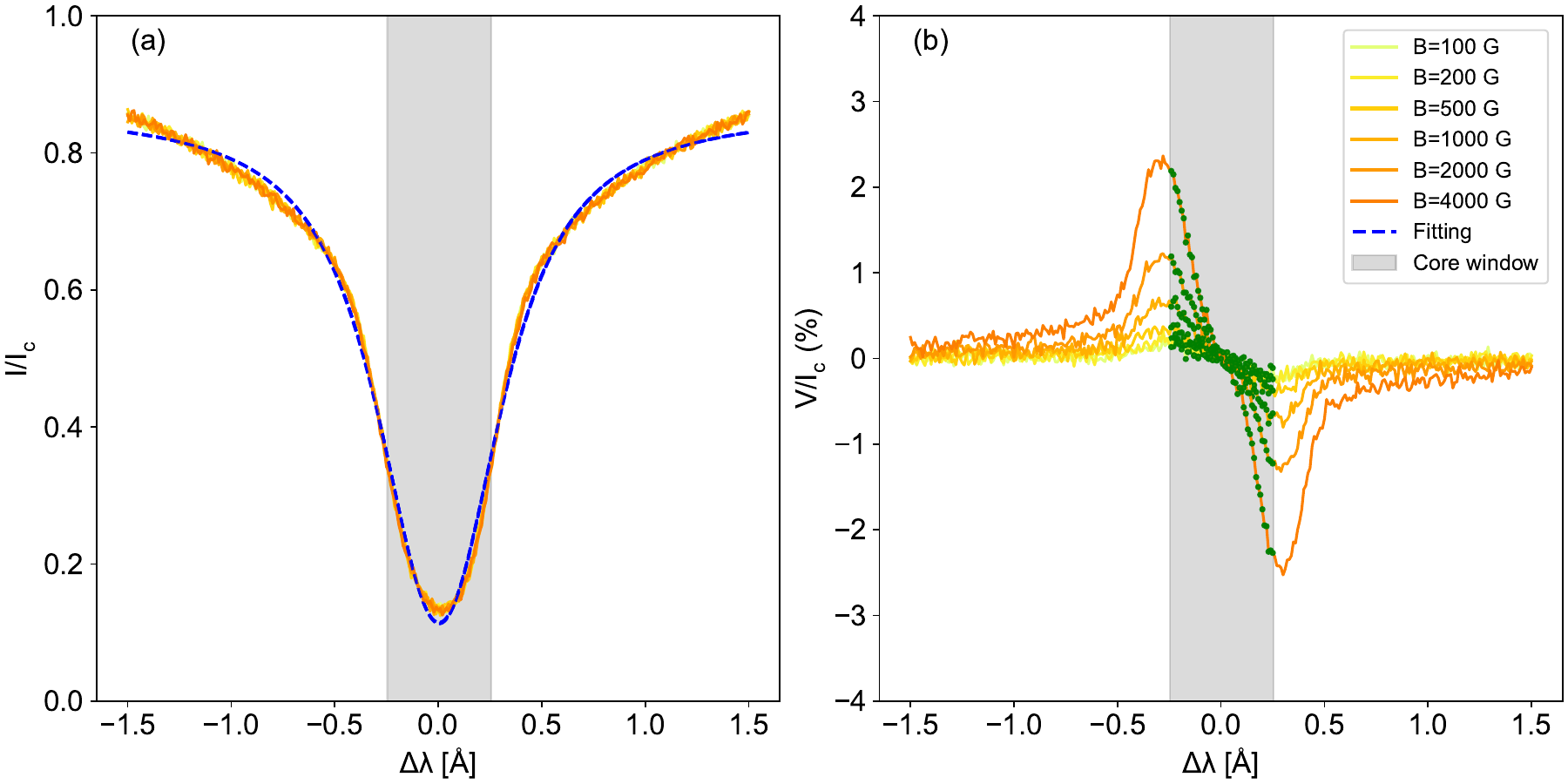}
		\caption{Synthetic Stokes profiles with different magnetic field strengths. Panel (a) shows Stokes I profiles (yellow) in the range from 100 -- 4000 G and the fitting curves (blue dashed) with Voigt profile. Panel (b) shows Stokes V/I$\mathrm{_c}$, with green dots (as well as gray span in (a) and (b)) indicating wavelength range of [$-$0.26, 0.26] \AA used for the inference of chromospheric magnetic fields using the WFA.}
		\label{fig:f1}
	\end{figure}

	One limitation of the WFA is that it assumes the atmospheric parameters, including the magnetic field components, LOS velocity, and Doppler width, remain approximately constant with optical depth. Strong gradients or abrupt variations with height may violate these assumptions and reduce the accuracy of the approximation. Circular polarization signals are generally stronger in narrow spectral lines than in broad ones, while scattering processes can dominate the linear polarization signal in weak-field atmospheres, particularly near the line center. 
	In addition, because the WFA expressions for Stokes Q and U depend on the second derivative of the intensity profile, the transverse magnetic field inferred from linear polarization is sensitive to line-broadening mechanisms, including microturbulence. Consequently, reliable inference of the transverse field under the WFA is only valid when Doppler shifts and line broadening remain relatively uniform within the line-forming region.
	Figure \ref{fig:f1} shows examples of Stokes I (left panel) and V/I$\mathrm{_c}$ (right panel) profiles with different magnetic field strengths. The shaded gray areas represent the wavelength range used for the inference of chromospheric magnetic fields using the WFA.
	
	\subsubsection{Non-LTE inversion: DeSIRe} \label{subsec:desire}
	
	\begin{figure*}[ht!]
		\centering
		\includegraphics[width=\linewidth]{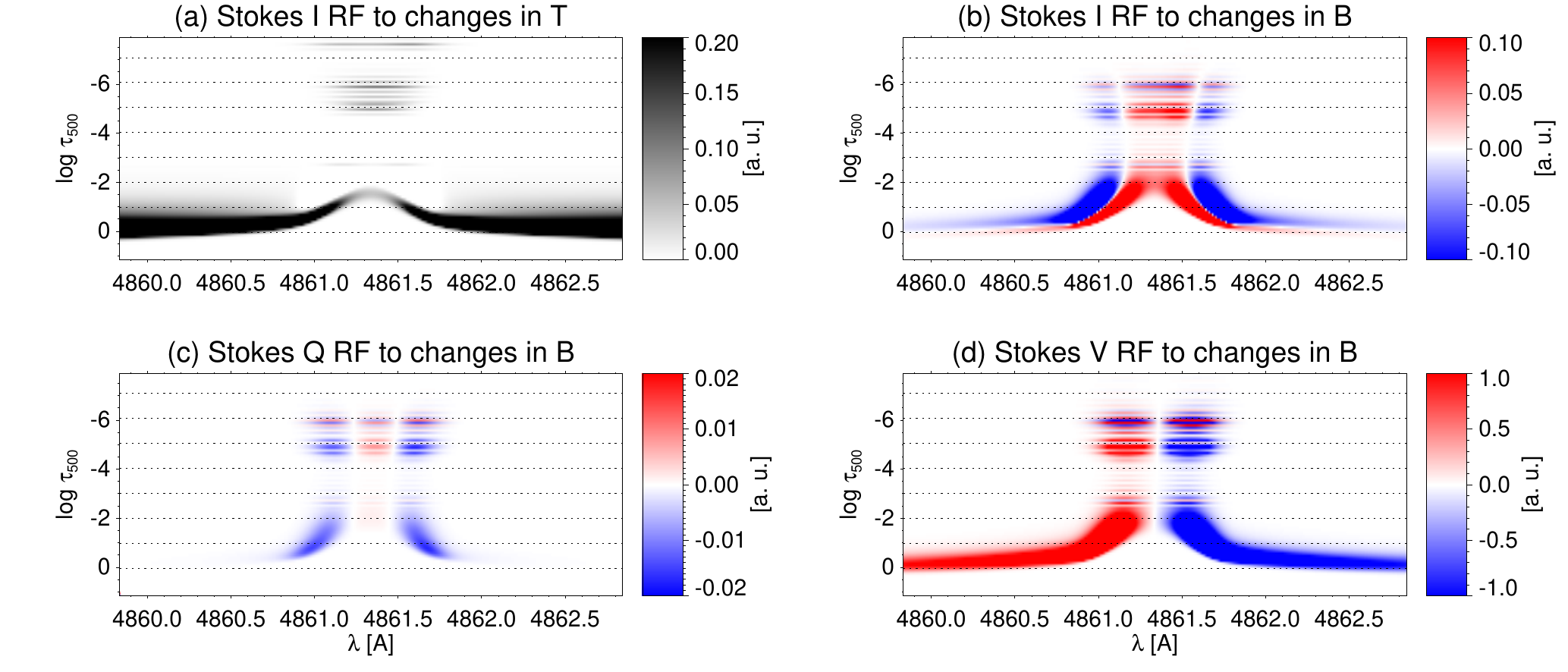}
		\caption{Response function maps of Stokes profiles to the change of temperature and magnetic field strength. (a) and (b) show response functions of Stokes I to T and B, respectively, along optical depth. (c) and (d) show response functions of Stokes Q and V to B, respectively. The maps are scaled to arbitrary unit for RF to temperature and magnetic field, separately . The dashed horizontal lines indicates optical depths \logt= [0, $-$7] with even interval $\Delta$\logt= $-$1.0. }
		\label{fig:f2}
	\end{figure*}
	
	\cite{2022A&A...660A..37R} introduced DeSIRe (Departure coefficient aided Stokes Inversion based on Response functions), a novel inversion code designed for fast and accurate analysis of spectropolarimetric data involving chromospheric lines that form under non-LTE conditions. DeSIRe integrates the SIR inversion code (for LTE conditions) with the RH radiative transfer solver (for non-LTE computations), using fixed departure coefficients to modify the analytical response functions (RFs) of LTE. This hybrid approach drastically reduces the number of iterations of full RTE solver needed during the inversion. DeSIRe solves the SE and RTE iteratively, optimizing the agreement between observed and synthetic spectra using a $\chi^{2}$ minimization scheme. Several atmospheric parameters are iteratively perturbed using RFs and nodes at different optical depths, with performance validated against three dimensional MHD simulations. DeSIRe provides a robust and publicly available tool for the solar physics community to analyze upcoming high-resolution chromospheric observations.
	
	Figure \ref{fig:f2} presents the RFs of the Stokes profiles in \hb~to temperature (T) and magnetic field strength (B) in solving RTE problems in non-LTE chromosphere. The RFs of Stokes I profiles to T and B show double peaks at \logt$\simeq$ $-$1 and \logt$\simeq$ $-$5.0, respectively (as shown in Figure \ref{fig:f2}(a) and (b)). Figure \ref{fig:f2}(c) and (d) show RFs of linear (Stokes Q at azimuth $\phi = 45^\circ$ and circular (Stokes V) polarization to the change of magnetic field strength ($\Delta$B = 1~G). It is clear that there are peaks of RF in chromospheric height at \logt$\simeq$ $-$5.0, and complex signals of RF is observed at  \logt $>$ $-$6.0 with mixed polarities. That indicated high noise due to inconsistent departure coefficient (DC) values given in DeSIRe with the non-LTE condition of the atmosphere model. The peaks in RF implicate the height of the solar atmosphere whose \hb\ polarized spectral intensity is the most sensitive to temperature and magnetic field change in the formal solutions of RTE.
	
	\subsection{Magnetic Field Configuration}
	To generate synthetic Stokes profiles, we adopted two magnetic configurations in the atmospheric models: constant vertical magnetic fields and linear gradient magnetic fields. Both configurations are created with a given boundary field strengths at the solar surface (\logt = 0).
	The boundary magnetic field strengths consisting of 100 values in the range of B = [0, 5000]~G are divided into 5 subranges, with 4 values in the range of [0, 100]~G, 40 values in the range of [100, 1000]~G, 20 values in the range of [1000, 2000]~G, 20 values in the range of [2000, 3500]~G, and 6 values in the range of [3500, 5000]~G.
	All values are uniformly distributed in each subrange so we have more data in the range of [100, 1000]~G, which is close to real chromospheric field characteristics. Five fixed values (15, 30, 45, 60, 75)$^\circ$ of the inclination angle are set for each boundary magnetic field, constructing 500 different models in each magnetic configuration.  In Sections \ref{sec:4.1} and \ref{sec:4.2}, the magnetic field strengths are set to constant along height. In Sections \ref{sec:4.3} and \ref{sec:4.4}, the magnetic field strengths are set to be linearly decreased in the range \logt =  [0,$-$7.0] and vanish at \logt = $-$7.0. 
	
	\section{Results}\label{sec:results}
	
	We assessed the reliability of WFA under typical chromospheric conditions (Section \ref{sec:4.1}) and used WFA-derived field estimates to initialize the non-LTE inversion procedure based on the DeSIRe code \citep{2022A&A...660A..37R} (Sections \ref{sec:4.2} -- \ref{sec:4.4}). This hybrid approach leverages departure coefficients to account for non-LTE deviations while maintaining computational efficiency.

	\subsection{Results from the WFA}\label{sec:4.1}
	
	\begin{figure}[ht!]
		\centering
		\includegraphics[width=0.9\linewidth]{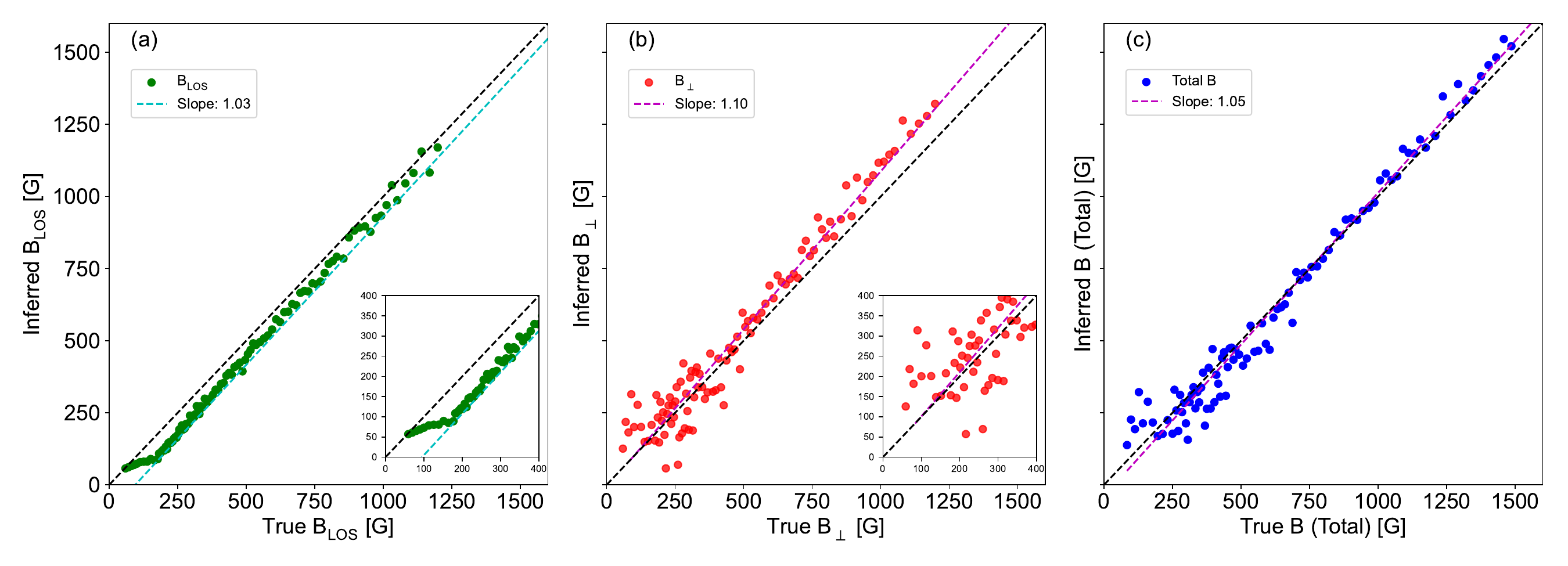}
		\caption{Magnetic field retrieved with WFA in the chromosphere at \logt= $-$5.0. 
		Inferred LOS components, transverse components, and calculated total field are plotted as a function of the model field strength in panel (a), (b), and (c), respectively. The inset in the lower right shows inferred fields in the weak field range [0, 400]~G using the same color scheme. The black dashed line indicates linear correlation expectation of 1.0 between inferred total (vertical/transverse) field with model field strengths. } 
		\label{fig:wfa}
	\end{figure}
	
	Figure \ref{fig:wfa} shows the magnetic field estimation using WFA approach under the magnetic configuration in which field strength linearly decreases with height, and is 45$^\circ$ inclined. We focus on the chromospheric magnetic field, of which the contribution function of emergent radiation has secondary peaks at $\pm$0.12~\AA. Thus, integration of the spectral profile only covers the wavelength range of $-$0.26 to 0.26~\AA\ in Stokes V profiles as indicated in Figure \ref{fig:f1} (b) to avoid blend contribution of photosphere and chromosphere. 
	Figure~\ref{fig:wfa}(a) and (b) show LOS and transverse components of field estimated with WFA, respectively, with results in weak field regime (0 -- 400~G) presented in corresponding zoomed panels.
	Eq.~(\ref{eq:1}) provides the LOS component of the magnetic field B$_{LOS}$. The total field strength is subsequently reconstructed by combining the LOS and transverse components derived from Eq.~(\ref{eq:6}) (see Figure~\ref{fig:wfa}(c)).
	It shows high correlation (least squares coefficient of 1.05) with model magnetic field inputs in the synthesis profile, the failed exception is obtained where the field strength is below 250~G. 
	With inclination angle of 45$^\circ$, the linear correlation coefficient of the LOS and transverse components with the total field strength is 0.71. While the inferred LOS fields are highly correlated with expected trend, showing a correlation coefficient of 0.73 (corresponding to 1.03 in linear correlation between inferred and true LOS components), there is a systematic bias in the mean (60~G).
	
	Scattering processes are sensitive to temperature fluctuations \citep{2002ApJ...565.1312U, 2012ApJ...749..136L, 2023ApJ...951..151C} and can contribute significantly to linear polarization in spectral lines, particularly near the line center  \citep{2001ASPC..236..161T, :2004aa}. This contribution can complicate the interpretation of transverse magnetic fields inferred from linear polarization.
	Therefore, Eqs.~(\ref{eq:4}) and (\ref{eq:5}) are  better suited in WFA calculation for transverse component of magnetic field by avoiding \hb\ line center.  Because the linear polarization of \hb\ is weak (less than1\% of I$_{c}$), the addition of noise to the profiles leads to larger discrepancies in the inferred transverse magnetic field.
	The results satisfied a least squares linear regression with slope of 0.77 (corresponding to 1.10 in linear correlation between inferred and true transverse components). The linear relation is obtained with the transverse components in the range of [400, 1200]~G. The retrieved values are constant at 250~G when transverse field is weaker than 250~G, resulting in the inferred magnetic field strength to be stronger with larger inclination and significantly deviates from linear relation to model field below 250~G. 
	
	\subsection{Constant magnetic fields}\label{sec:4.2}
	\begin{figure*}[ht!]
		\centering
		\includegraphics[width=\linewidth]{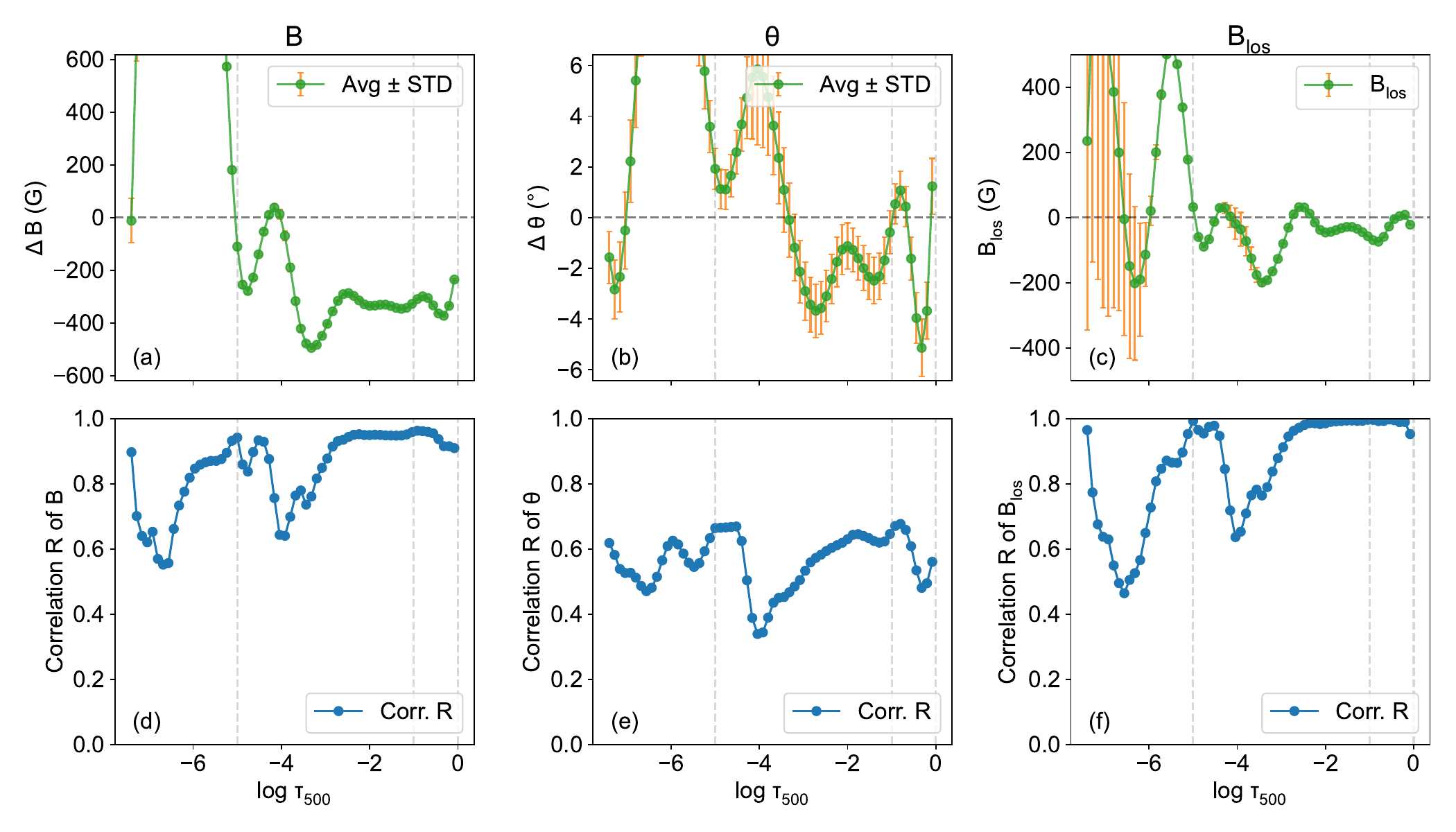}
		\caption{1D FAL-C model atmosphere of quiet Sun with constant magnetic field strength along height. (a)--(d) show variation of difference between inversion retrieved values and model for total field strength, inclination angle, and LOS field, respectively. Green dotted curves indicate average of differences from all inversion sets vs optical depth. Yellow bars mark standard deviation of differences between retrieved values and model. (e)--(h) show Pearson's correlation coefficients of inversion retrieved vs. model for corresponding values at top panels. The vertical dashed lines indicate optical depths at \logt= $-$5.0, $-$1.0, and 0. }
		\label{fig:f5}
	\end{figure*}

	\begin{figure*}[ht!]
		\centering
		\includegraphics[width=0.8\linewidth]{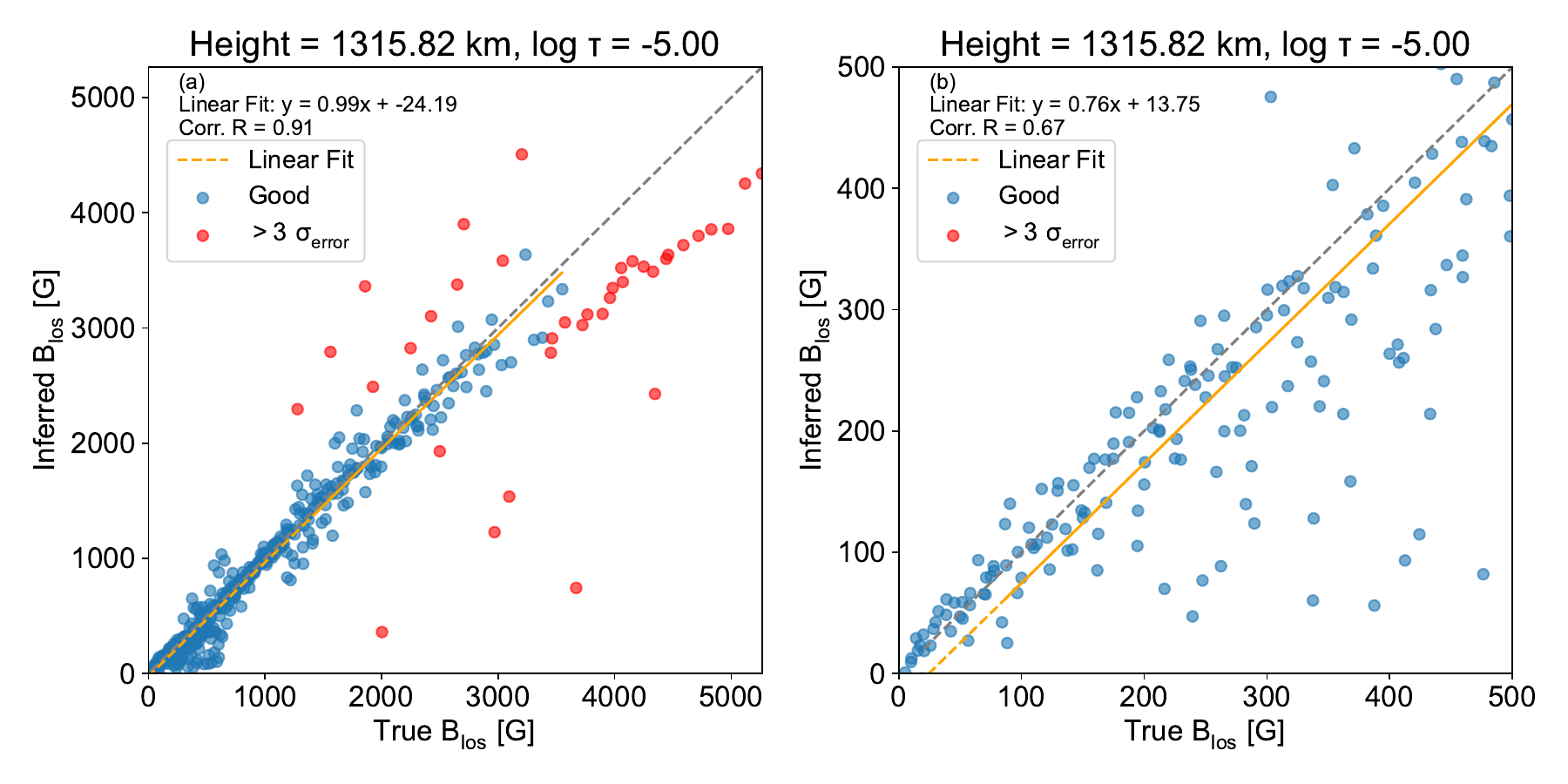}
		\vspace{0.3cm}  
		\includegraphics[width=0.8\linewidth]{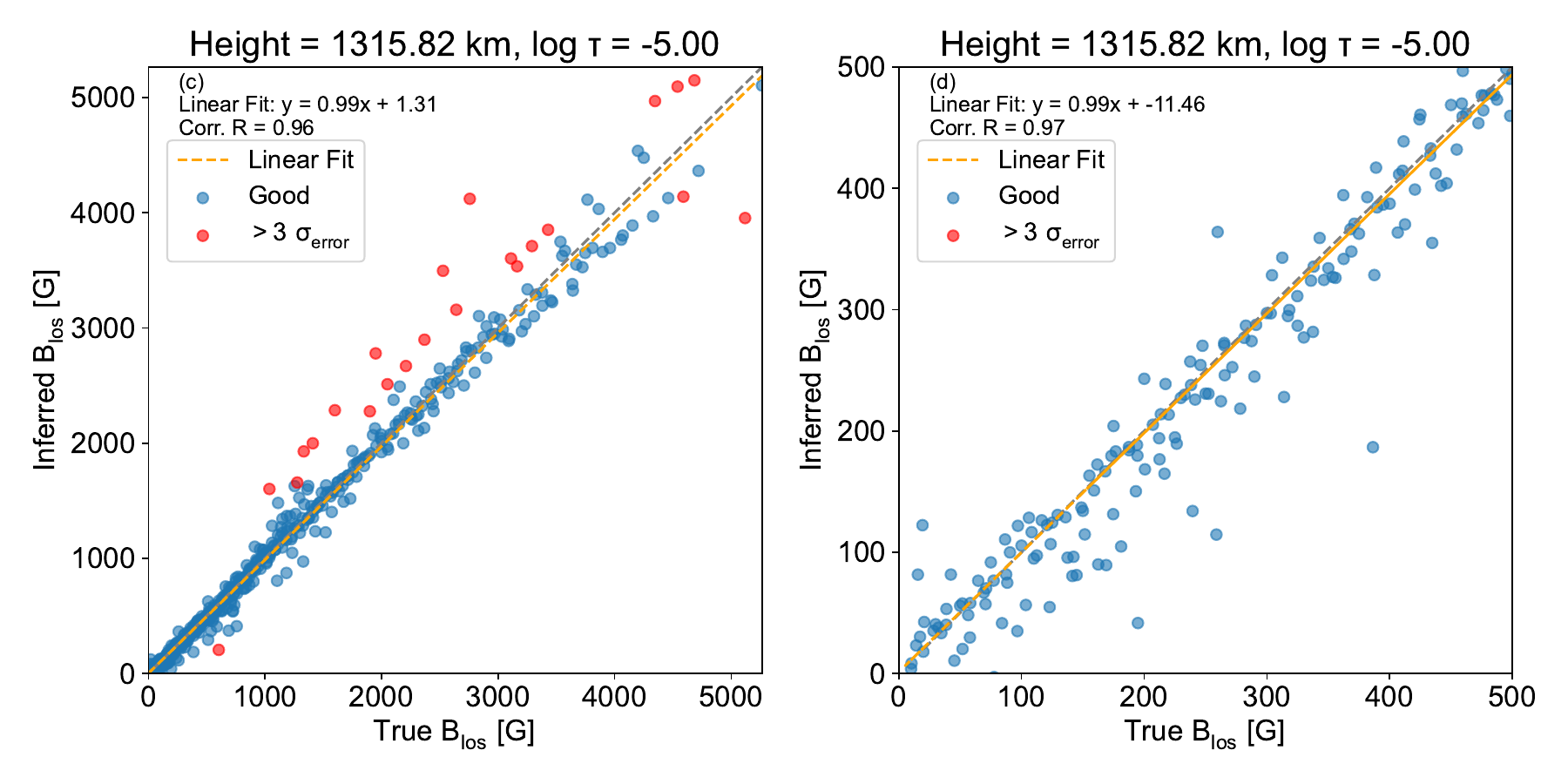}
		\caption{Inference of LOS magnetic field vs constant field strength model. Panels (a) and (b) show the inversion results run in full magnetic field range and weak regime (0 to 500~G), respectively, with fixed initial guess B=1000~G and $\theta$=30$^\circ$. Gray dashed indicate the line of equality. Yellow dashed lines indicate linear fitting line. Red dots indicate outliers of fitting results that deviate from the linear fit by 3$\sigma$. Panels (c) and (d) show the inversion results conducted with WFA derived initial guess for B and $\theta$ using the same plotting convention. }
		\label{fig:f6}
	\end{figure*}

	By adapting a weight factor of 5 to the Stokes profiles at the line core wavelength range, which is determined by effective Doppler width and valid for magnetic measurement with WFA, chromospheric contribution is enhanced in non-LTE inversion when utilizing DeSIRe to the spectrum from [$-$1.5, 1.5]~\AA\ offset the line center. 	The effective Doppler width $\sigma_w$ = $\sqrt{2\pi}\sigma_v$ is estimated from a Voigt profile fit, where $\sigma_v$ represents the standard deviation of the Gaussian component corresponding to Doppler broadening. This parameter provides a characteristic measure of the line width but does not fully account for additional broadening mechanisms such as Stark broadening or instrumental effects. Therefore, the spectral fitting window used in the inversion is chosen to encompass the chromospheric sensitivity range of the \hb\ line rather than being defined solely by the Voigt profile width (Sections~\ref{sec:4.3}--\ref{sec:4.4}).
	The first set of tests was conducted using a series of constant magnetic field strengths within a quiet-Sun semi-empirical atmosphere, based on the FAL-C model. Magnetic field inversions were performed in two runs: in the first run, the inversion used fixed initial guesses for both the field strength and inclination angle; in the second run, the initial values were derived from the Weak Field Approximation (WFA) estimates of magnetic field strength and inclination.
	
	We evaluated the dependence of inversion accuracy on the formation height. Figure \ref{fig:f5} shows the correlation between inferred magnetic field information and that of different optical depths in the 1D simulation. The results are obtained by using non-LTE inversion with WFA results as initial guess. Green curves in the top panels show difference between inferred and model magnetic field strength, inclination angle, and LOS component of field, respectively. The orange bars represent root-mean-square (r.m.s) error. The blue curves in the bottom panels show corresponding Pearson correlation coefficients (c.c.). The vertical dashed lines indicate heights at which \logt= $-$5.0, $-$1.0, and 0. The highest spatial correlation of magnetic field with model is obtained at \logt= $-$1.2 and the second peak appeared at \logt= $-$5.0 (see Figure \ref{fig:f5}(b)). 
	The largest disagreement between inferred field and model occurs at locations where~\logt~$\le$ $-$6.0. This aligns with the findings in Figure \ref{fig:f2}. 
	The inclination angle also shows two peaks in spatial correlation but the correlation coefficient is low (0.67 at the primary peak), meaning the inferred transverse fields have larger uncertainty while the total fields agree with model (Pearson cc.  = 0.95 at the primary peak). This is consistent with the conclusions obtained through WFA in Figure 5. Due to the weak signal of linear polarization and its absence from the current measurement targets of the SFMM, in the following sections of this work, we focus on the results of the LOS field.

	Figure \ref{fig:f6} presents a comparison between the inferred and simulated magnetic fields. The top panels show the correlation between the LOS component of the magnetic field inferred from inversions with fixed initial guesses and the corresponding LOS components in the model, Figure \ref{fig:f6}(b) show detailed correlation in field strength below 500~G. The bottom panels illustrate the same comparison, but using WFA-derived values as the initial guess in the inversion process.
	Red dots in each panel represent data points that deviate by more than 3$\sigma$ from the linear regression, based on the r.m.s error of the fit, highlighting cases with significant discrepancy between the inferred and input magnetic fields.
	An improvement in the coherence between the magnetic fields inferred using the WFA results as initial guess and the magnetic fields in the model is evident in Figure \ref{fig:f6}. The linear regression coefficient and Pearson correlation approaches unity in Figure \ref{fig:f6}(c) and (d), indicating better agreement when using WFA-derived initial guesses. Notably, the number of outliers in the inversion results is significantly reduced within the magnetic field strength ranges of [0, 500]~G (Figure \ref{fig:f6}(b) and (d)). This suggests that adopting physically informed initial conditions, such as those from the WFA, can improve the accuracy and robustness of inversion outputs, particularly for weak to moderate field strengths.
	
	\subsection{LOS Magnetic field with linear gradient in FAL-C Quiet Sun Model Atmosphere}\label{sec:4.3}
	
	\begin{figure*}[ht!]
		\centering
		\includegraphics[width=0.8\linewidth]{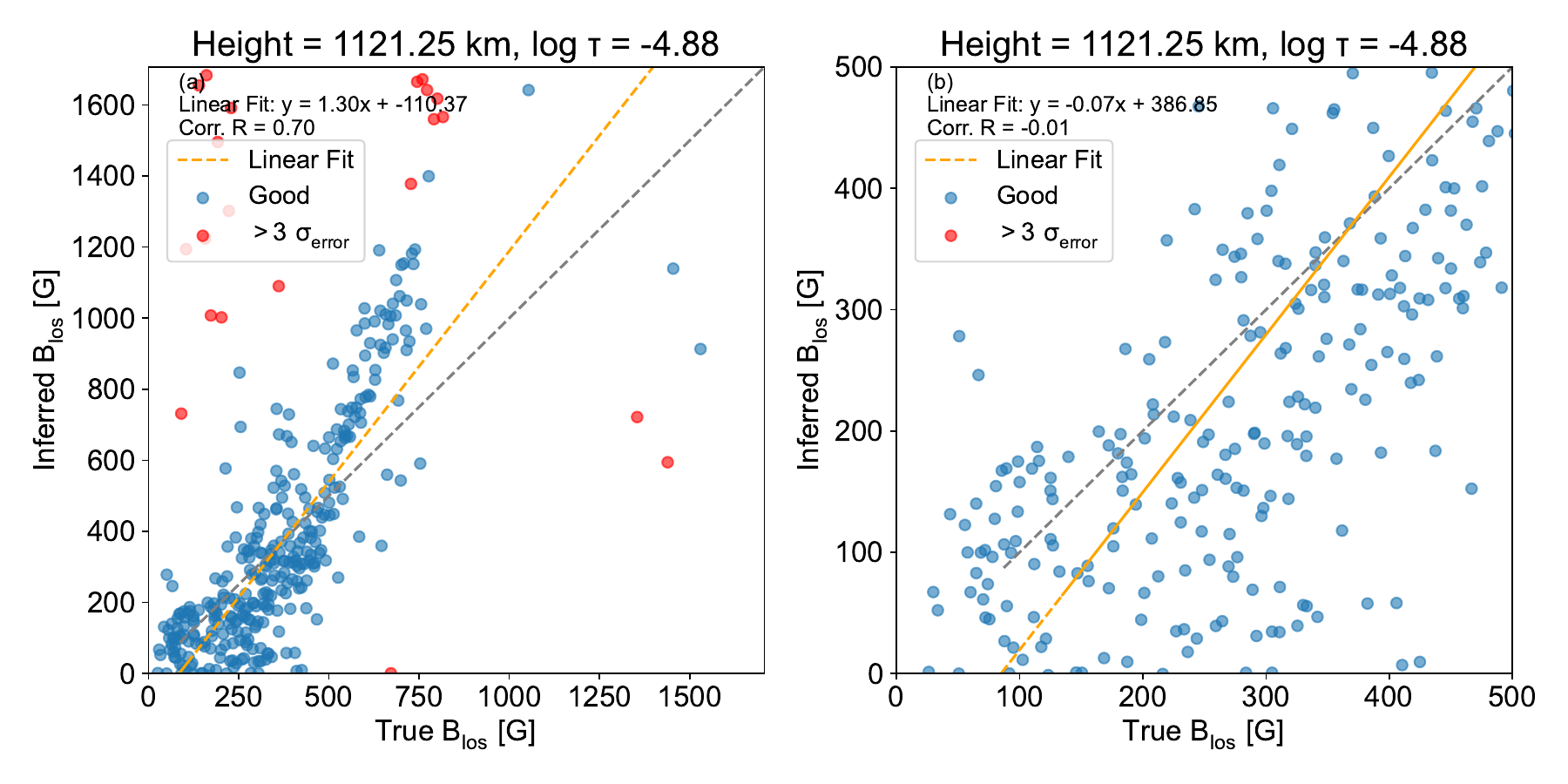}
		\vspace{0.3cm}  
		\includegraphics[width=0.8\linewidth]{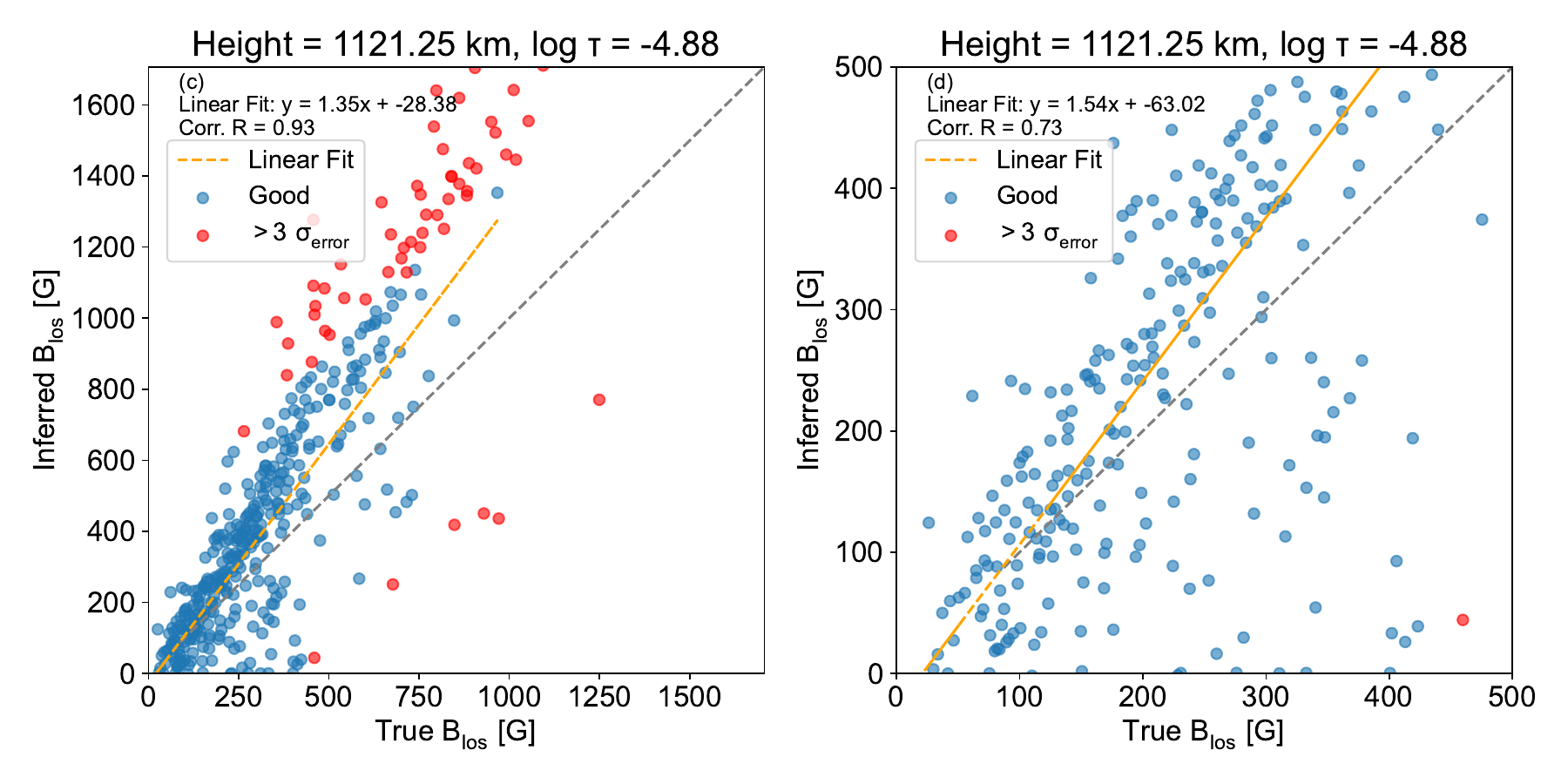}
		\caption{The same with Figure \ref{fig:f6} but for magnetic field with linear gradients.}
		\label{fig:f8}
	\end{figure*}
	
	In the next set of simulations, we prescribed a magnetic field strength that decreases linearly with height in the solar atmosphere and repeated the inversion procedures described in Section \ref{sec:4.2}.
	Figure \ref{fig:f8} presents the results for the inferred LOS component of the magnetic field, using the same wavelength range selected based on Voigt profile weighting. As shown in Figure \ref{fig:f8} (a) and (b) display the linear correlation between the inferred and simulated LOS magnetic fields obtained with fixed initial guesses, while Figure \ref{fig:f8} (c) and (d) show the corresponding results using WFA-derived initial guesses.     
	In contrast, although the WFA-derived initial guess improves the linear correlation between the inferred and simulated magnetic fields (Figure \ref{fig:f8} (c) and (d)), the linear regression coefficient still deviates from unity. Specifically, the inferred values of the LOS magnetic field are systematically overestimated by approximately 24--61\%, depending on the inclination angle and field strength range. This indicates that while the use of WFA-based initialization enhances the overall trend, it does not fully resolve the bias introduced by the height-dependent field gradient. 
	
	\begin{figure*}[ht!]
		\centering
		\includegraphics[width=0.8\linewidth]{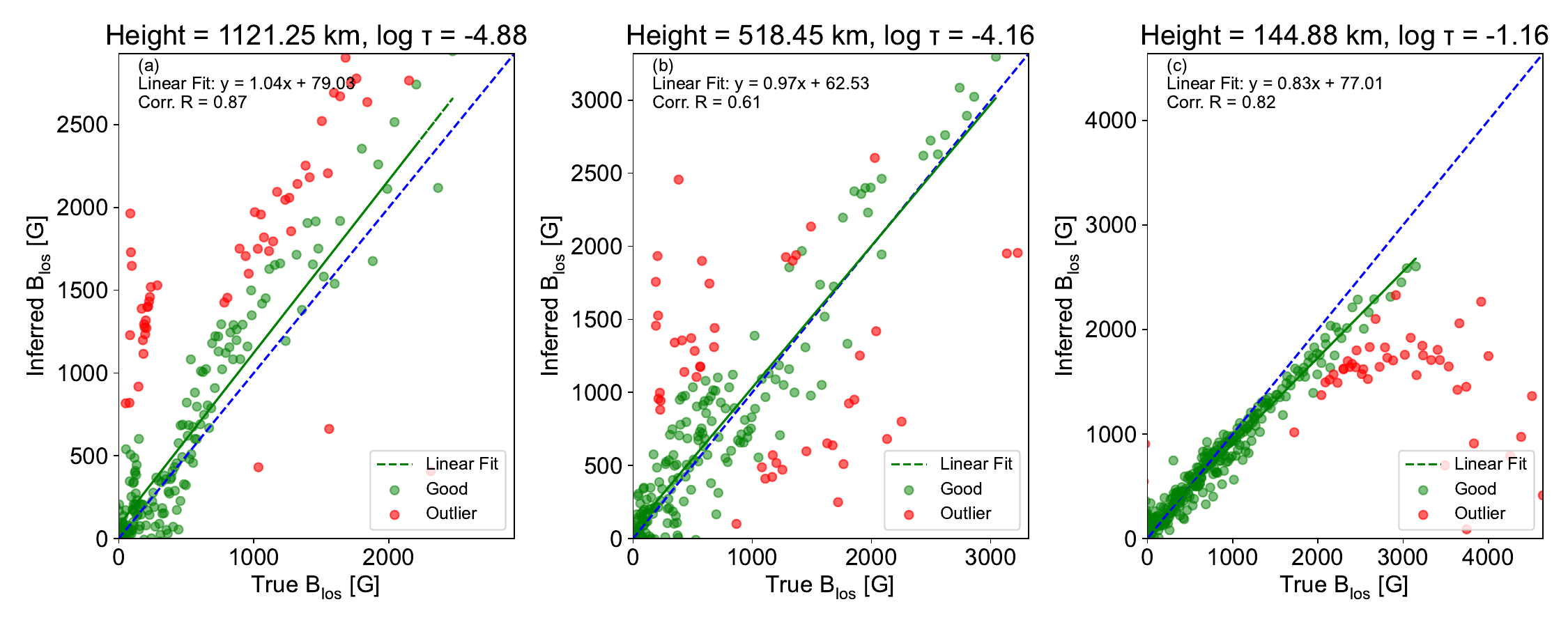}
		\vspace{0.3cm}  
		\includegraphics[width=0.8\linewidth]{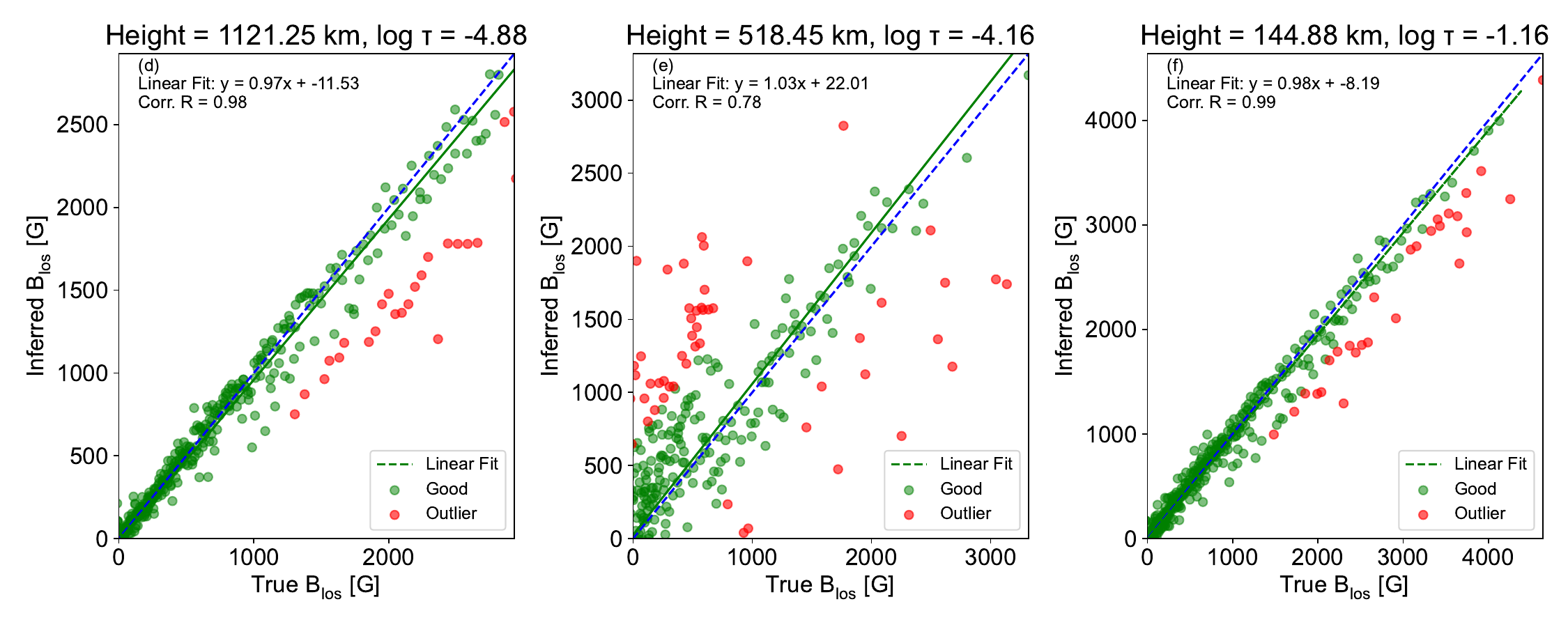}
		\caption{Inferred LOS field strength using DeSIRe vs model field at different heights. The model magnetic field strength has linear gradient along height. (a)--(c) show the inversion results run with fixed initial guess B = 1000~G and $\theta$ = 30$^\circ$. Blue dashes indicate the line of equality. Green dashed lines indicate linear fitting line. Red dots indicate outliers of fitting results that deviate from the linear fit by 3$\sigma$. Green dots indicate good inversion results that satisfy the linear fit. (d)--(f) show the inversion results conducted with WFA derived initial guess for B and $\theta$ using the same plotting convention.}
		\label{fig:f7}
	\end{figure*}
	
	Upon reviewing the synthetic spectral profiles generated under both constant magnetic field and linearly decreasing field configurations, we observed that the Doppler width in the quiet Sun atmosphere varied only marginally, remaining within the range of [0.24, 0.26]~\AA. Based on this consistency, we fixed the weighted wavelength range for the inversion to [$-$0.13, 0.13]~\AA~relative to the line center, and repeated the inversion tests using the linear gradient magnetic field configuration.
	Figure \ref{fig:f7} presents the results for the inferred LOS magnetic field components at three different heights: 1121~km, 518~km, and 145~km (corresponding to \logt =$-$4.88, $-$4.16, $-$1.16), representing height of response function peaks in chromosphere and photosphere, and an intermediate height. In comparison with Figure \ref{fig:f8}, the correlation between the inferred and simulated magnetic fields reaches unity at a height of 1121 km, while a relatively high degree of coherence is also obtained at 145~km. These results are consistent with the predicted formation height (1100--1350~km) of the \hb\ line in the chromosphere as described by the FAL-C model in Figure \ref{fig:cf}.

	\subsection{LOS Magnetic field with linear gradient in FAL-S Sunspot Model Atmosphere }\label{sec:4.4}
	
	\begin{figure*}[ht!]
		\centering
		\includegraphics[width=0.8\linewidth]{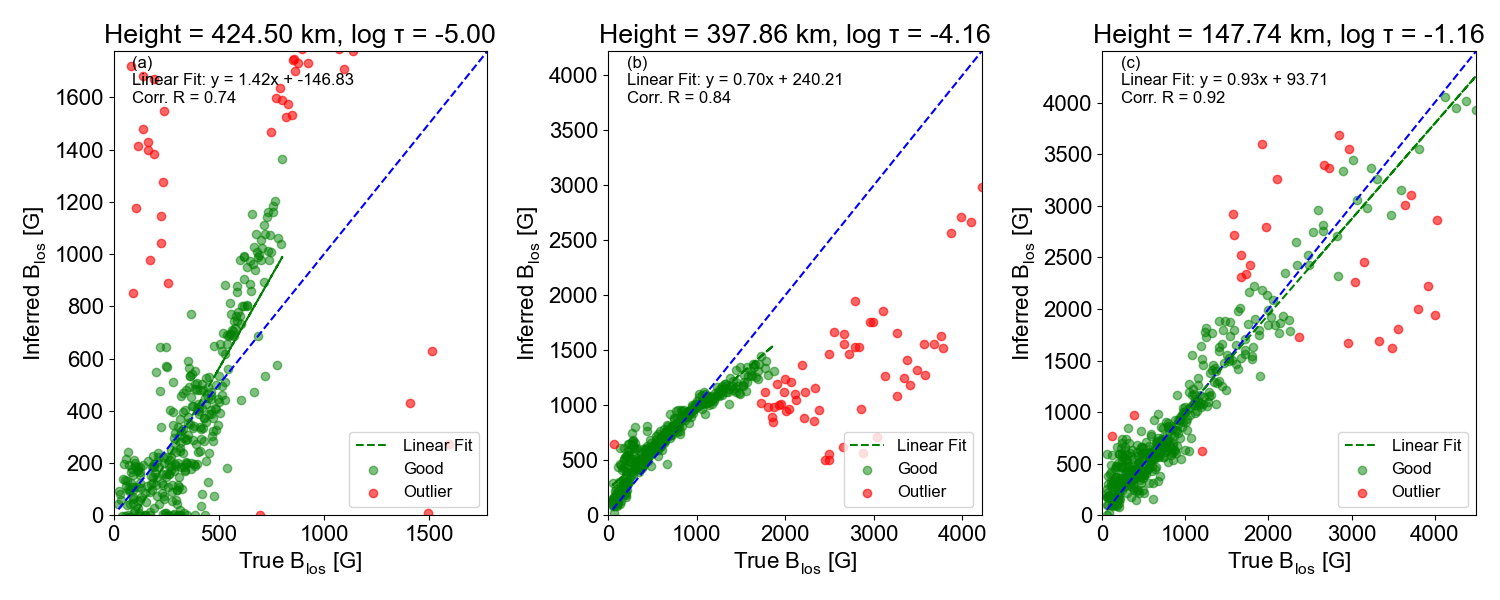}
		\vspace{0.3cm}  
		\includegraphics[width=0.8\linewidth]{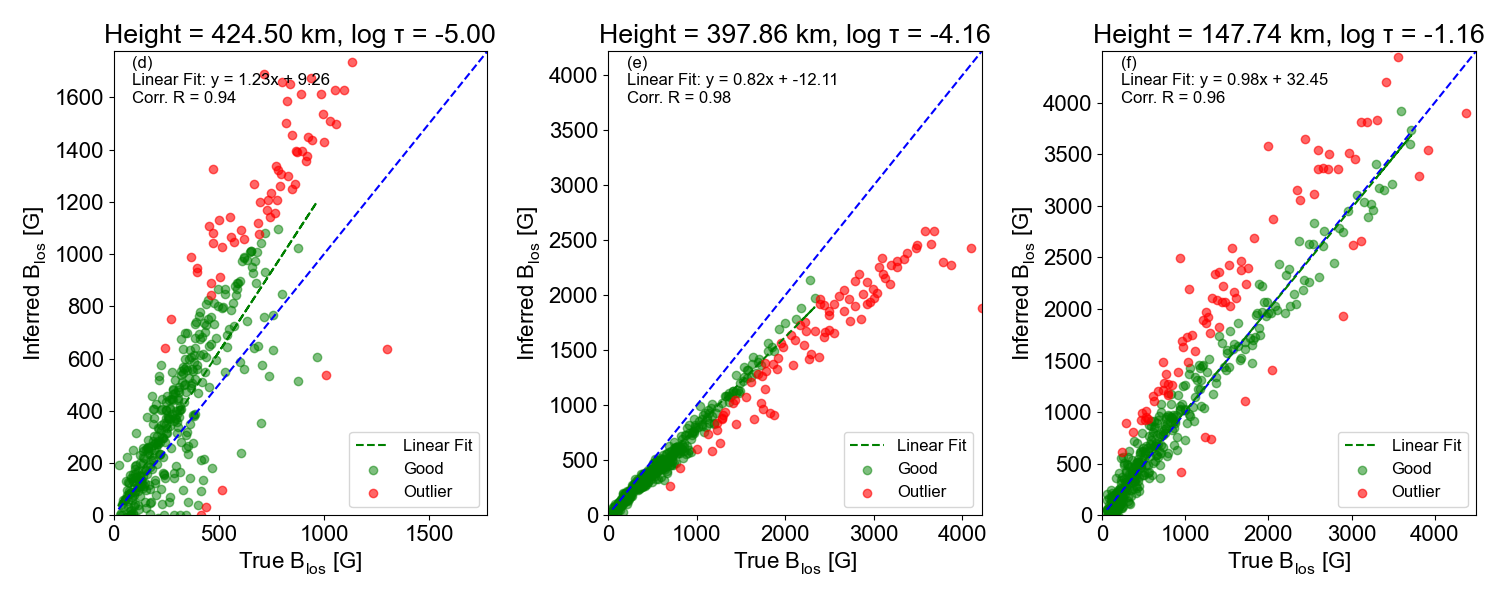}
		\caption{The same with Figure \ref{fig:f7} but for sunspot model.}
		\label{fig:f9}
	\end{figure*}

	In the sunspot model atmosphere, it is reasonable to assume a linearly decreasing magnetic field gradient in the umbral chromosphere, consistent with magnetic flux conservation and the natural decline in field strength from the umbra to the penumbra. For this case, inversions were performed on synthetic Stokes profiles generated with the FAL-S model. The formation height of \hb\ depends strongly on the local thermal stratification: in the cooler umbral chromosphere, the $\tau$= 1 surface forms deeper in the photosphere than in the quiet Sun. Consequently, the contribution function peaks in the photosphere, where the source function is higher, leading to stronger photospheric dominance.
	
	Figure \ref{fig:f9} presents the inversion results at different geometric heights in the sunspot atmosphere. Using Pearson’s correlation coefficient (R) as a metric for inversion accuracy, the best linear correlation between the inferred and model magnetic fields is obtained at 147~km, when WFA-derived values are used as initial guesses. This height corresponds to \logt = $-$1.16 (Figure \ref{fig:f9} (e)), which is approximately the same as the primary formation height in the quiet Sun photosphere (see Section \ref{sec:4.3}). 
	The R value attains its highest value at the height of 398 km, which is corresponding to optical depth \logt = $-$4.16 with lower temperature than QS chromosphere, whereas the linear slope remains different from unity. The inferred field strength is underestimated by \sm18\%, with deviations exceeding $3\sigma$ for LOS fields stronger than 2000~G. Magnetic field strength in the chromosphere are typically in the range of 100 -- 1000~G, in which the results show good correlation in the chromospheric (photospheric) heights (\logt = $-$4.16 in Figure \ref{fig:f9} (e) and \logt = $-$1.16 in Figure \ref{fig:f9} (f)). 
	
	To compare inversion accuracy under different magnetic configurations, Pearson’s correlation coefficients were calculated for both the quiet-Sun and sunspot models. As summarized in Table \ref{tab:tab1}, the highest correlation values are consistently obtained when height variations of the magnetic field are excluded. However, the sunspot model yields systematically lower correlation coefficients, indicating more significant inversion outliers. These results demonstrate that introducing a linear magnetic gradient into the inversion setup increases the overall uncertainty, particularly in strongly magnetized atmospheres such as sunspots.

	\begin{table}[ht!]
		\centering
		\begin{tabular}{|ccccc|}
			\hline
			Pearson's R & \multicolumn{2}{c|}{Constant $\mathbf{B}$}& \multicolumn{2}{c|}{$\mathbf{B}$ with linear gradient} \\
			\hline
			Field (G) & (0--5000) & (0--1000) & (0--5000) & (0--1000)\\
			\hline
			FAL-C & 0.96$\pm$0.02 & 0.98$\pm$0.01 & 0.87$\pm$0.09 & 0.89$\pm$0.05\\
			FAL-S & 0.87$\pm$0.05 & 0.82$\pm$0.09 & 0.71$\pm$0.13  & 0.77$\pm$0.15\\
			\hline
		\end{tabular}
		\caption{Pearson's correlation coefficients of inferred field vs model field for quiet Sun (FAL-C) and sunspot (FAL-S). The model fields are constructed with constant $\mathbf{B}$, and linearly varying $\mathbf{B}$ with height at different inclination angles.}
		\label{tab:tab1}
	\end{table}

	\section{Conclusions and Discussions}
	
	In this work, we investigated the feasibility of using the \hb~4861~\AA\ spectral line as a diagnostic tool for chromospheric magnetic fields by applying non-LTE inversion techniques and comparing them with the WFA. Synthetic Stokes profiles were generated using semi-empirical FAL atmospheric models under both constant and linearly decreasing magnetic field configurations, spanning field strengths from 0 to 5000 G.
	
	Our main results are summarized as follows:
	\begin{enumerate}
		\item Formation height sensitivity: The \hb\ line exhibits dual sensitivity, with contributions from both the photosphere (\sm400 km) and the middle chromosphere (\sm1000--1500 km), consistent with previous studies. Inversions performed at ~1315 km showed the best agreement with quiet Sun model inputs, aligning with predicted chromospheric formation heights.
		\item In the quiet Sun: WFA-derived initial guesses significantly improved the coherence between inferred and simulated fields. Outliers were notably reduced in the field ranges 250--500~G and 2000--4000~G, demonstrating that physically motivated initial conditions can enhance inversion robustness.
		\item In sunspot: The \hb\ formation height shifts \sm900~km deeper than in the quiet Sun, leading to stronger photospheric dominance. The best correlation is achieved at \sm397~km using WFA-derived initial guesses, though field strengths are underestimated by \sm18\% and inversion uncertainties remain higher in strongly magnetized atmospheres.
		\item Instrumental considerations: Synthetic tests incorporating instrumental broadening (\sm85 m\AA) and realistic noise levels showed that polarization signals remain measurable under quiet-Sun conditions, suggesting feasibility of applying this method to full-disk \hb\ observations from instruments such as the Solar Full-Disk Multi-layer Magnetograph at Ganyu Solar station.
	\end{enumerate}
	
	Overall, our results validate the effectiveness of non-LTE inversions of the \hb\ line for retrieving chromospheric magnetic fields, particularly the LOS component, across both weak and strong field regimes.
	
	Our results highlight several important aspects for advancing chromospheric magnetometry with hydrogen Balmer lines. The WFA remains a valuable and computationally efficient tool for providing initial conditions, especially in quiet-Sun regions where field strengths are below about 1200 G.  
	Nevertheless, its tendency to overestimate magnetic fields in the presence of vertical gradients underscores the importance of full non-LTE inversions for reliable diagnostics. Additionally, field estimates with WFA in fixed wavelength range provides better initial values in non-LTE solutions. 
	An adaptive hybrid approach, combining the speed of WFA with the accuracy of non-LTE solvers, may offer a promising path forward.
	
	The \hb\ line exhibits sensitivity to magnetic field in multiple atmospheric layers, with its strongest response appearing near 130~km at line wing (\>0.13~\AA) while the secondary coherence closer to the line core in the middle chromosphere around 1300 km. This dual sensitivity complicates the inversion process, as it is difficult to isolate contributions from different heights, but it also provides opportunities for multi-layer probing of chromospheric structure. In this context, the use of \hb\ in combination with other lines, such as \ha\ or Ca II, may provide more robust reconstructions of stratified magnetic fields.
	
	As emphasized by \citet{2025ApJ...987L..39M}, Balmer lines are excellent probes of dynamic processes but face challenges regarding magnetic sensitivity. The increased line width of \hb\ can be attributed to several factors, including thermal and nonthermal heating, turbulent motions, and magnetohydrodynamic waves. 
	The \hb\ spectral line requires higher temperature excitation, making umbral chromosphere generally transparent \citep{2020SCPMA..6319611Z}. Thus, the sunspot model with cooler temperature has lower formation height. 
	A notable outcome of previous analysis \citep{2025ApJ...988..232C} is that Kurucz (no chromosphere) and FALXCO (low chromospheric temperature) reproduce \hb\ cores better than hotter chromospheric models, despite earlier expectations \citep{2008ApJS..175..229A} that \hb\ should form substantially in the chromosphere, suggesting that overly hot chromospheric models (e.g., FAL-C) shift line formation too high, underestimating intensities. 
	
	Although the underlying atmospheric model was static and one-dimensional (FAL-C/FAL-S), we enabled dynamic recomputation of atomic level populations during each iteration of the inversion process using the RH code, when applying the DeSIRe algorithm. This allowed continuous updating of departure coefficients to reflect non-LTE conditions more accurately. We evaluated the inversion performance using three different approaches for defining the spectral fitting window: (1) the effective width derived from Voigt profile fitting, (2) the standard deviation from Gaussian fitting, and (3) a fixed spectral window of $\pm$130 m\AA\ (i.e., a total of 260 m\AA) centered on the line core.
	The sensitivity of spectral width to magnetic field strengths is negligible compared with Doppler velocity variation. For cases assuming a constant magnetic field and similar LOS velocity, both the fixed window and effective width methods yielded comparable inversion accuracy within the FAL-C model atmosphere. However, under conditions with spatially varying magnetic fields, the fixed spectral window approach resulted in more accurate inversion outcomes.
	
	The presence of magnetic fields introduces fine-scale structures that cannot be captured in one-dimensional or two-dimensional models. A realistic treatment therefore requires fully three-dimensional radiative MHD simulations, although these impose significant computational costs. While single-line inversions, such as those based solely on \hb\, can provide valuable information, they are inherently limited in their ability to separate chromospheric and photospheric contributions. 
	Since the emergent profiles can be reproduced by different atmospheric configurations, inversions are intrinsically degenerate, particularly in the presence of magnetic or thermal gradients. In our tests, this effect manifested as systematic overestimation of the LOS field strength in linear-gradient atmospheres, where the inversion could not fully disentangle contributions from overlapping layers. The reliance on initial conditions becomes stronger under such circumstances; using WFA-derived estimates improved convergence and reduced outliers, but uncertainties remain. These results underscore the need for multi-line inversions or additional observational constraints to mitigate scattering-induced ambiguities in chromospheric magnetic diagnostics.
	
	Applying this inversion framework to full-disk \hb\ observations offers an opportunity to obtain global maps of chromospheric magnetic fields, complementing existing diagnostics based on \ha\ and He I. Such observations are particularly valuable for understanding canopy fields, small-scale reconnection events, and flare-associated dynamics. The ability to capture both quiet-Sun and active-region fields also makes \hb\ an important bridge between local high-resolution studies with telescopes such as DKIST and SFMM, and global synoptic monitoring programs.
	
	The use of one-dimensional semi-empirical models restricts the ability to capture the inherently dynamic nature of the chromosphere, including shocks, fibrils, and transverse inhomogeneities. Discrepancies at optical depths log$\tau$= $-$6 suggest reduced sensitivity to the upper chromosphere. Future work extending this inversion framework to three-dimensional MHD models, such as Bifrost or MURaM, and applying it to multi-line spectropolarimetry would provide more realistic validation. Although our focus has been on the solar atmosphere, the methods developed here are potentially powerful for constraining magnetic parameters in a wider astrophysical context. Hydrogen Balmer lines are essential diagnostics in stellar atmospheres, where non-LTE effects, turbulence, and magnetic fields are equally important. Thus, the inversion techniques explored here may eventually be extended to stellar magnetometry and other astrophysical plasmas.
	
	\normalem
	\begin{acknowledgements}
		This work is supported by National Key R\&D Program of China No. 2021YFA1600500, National Natural Science Foundation of China (grant No. 12425301), Beijing Natural Science Foundation (grant NO. 1254055). 
		We gratefully acknowledge insightful discussions with Dr. Hui Tian (Peking University) and Dr. Hao Li (National Space Science Center), which contributed to the development of this work. 
		We also thank the anonymous reviewer for his/her constructive comments.
		We acknowledge the support from the Specialized Research Fund for State Key Laboratory of Solar Activity and Space Weather and the Chinese Meridian Project.The instrumental information for the chromospheric full-disk \hb\ filtergram was provided by Huairou Solar Observing Station, National Astronomical Observatories. 
	\end{acknowledgements}

	\bibliography{reference}
	\bibliographystyle{raa}
	
\end{document}